\documentclass[letterpaper,twocolumn,english,aps,pra,superscriptaddress,floatfix]{revtex4}
\usepackage{float}
\usepackage{amsmath}
\usepackage{graphicx}
\usepackage{amssymb}
\usepackage{esint}

\makeatletter

\providecommand{\tabularnewline}{\\}

\@ifundefined{textcolor}{}
{%
 \definecolor{BLACK}{gray}{0}
 \definecolor{WHITE}{gray}{1}
 \definecolor{RED}{rgb}{1,0,0}
 \definecolor{GREEN}{rgb}{0,1,0}
 \definecolor{BLUE}{rgb}{0,0,1}
 \definecolor{CYAN}{cmyk}{1,0,0,0}
 \definecolor{MAGENTA}{cmyk}{0,1,0,0}
 \definecolor{YELLOW}{cmyk}{0,0,1,0}
 }

\renewcommand{\[}{\begin{equation}}

\renewcommand{\]}{\end{equation}}

\makeatother

\usepackage{babel}

\makeatother

\usepackage{babel}

\makeatother

\usepackage{babel}

\makeatother

\usepackage{babel}

\begin{document}
\global\long\def\Cl{\mathrm{Cl}^{-}}

\global\long\def\Na{\mathrm{Na}^{+}}

\global\long\def\K{\mathrm{K}^{+}}

\global\long\def\Ca{\mathrm{Ca}^{2+}}

\global\long\def\Mg{\mathrm{Mg}^{2+}}

\global\long\def\eV{\,\mathrm{eV}}

\global\long\def\avg#1{\langle#1\rangle}

\global\long\def\p{\prime}

\global\long\def\dg{\dagger}

\global\long\def\ket#1{|#1\rangle}

\global\long\def\bra#1{\langle#1|}

\global\long\def\proj#1#2{|#1\rangle\langle#2|}

\global\long\def\inner#1#2{\langle#1|#2\rangle}

\global\long\def\tr{\mathrm{tr}}

\global\long\def\pd#1#2{\frac{\partial#1}{\partial#2}}

\global\long\def\spd#1#2{\frac{\partial^{2}#1}{\partial#2^{2}}}

\global\long\def\der#1#2{\frac{d#1}{d#2}}

\global\long\def\im{\imath}

\global\long\def\dI{\Delta I_{\mathrm{rel}}}

\global\long\def\erfc{\mathrm{erfc}}

\renewcommand{\onlinecite}[1]{\cite{#1}}

\title{Dehydration and ionic conductance quantization in nanopores}

\author{Michael Zwolak}

\affiliation{Theoretical Division, MS-B213, Los Alamos National Laboratory, Los
Alamos, NM 87545}

\author{James Wilson}

\affiliation{Department of Physics, University of California, San Diego, La Jolla,
CA 92093}

\author{Massimiliano Di Ventra}

\affiliation{Department of Physics, University of California, San Diego, La Jolla,
CA 92093}

\date{\today{}}
\begin{abstract}
There has been tremendous experimental progress in the last decade
in identifying the structure and function of biological pores (ion
channels) and fabricating synthetic pores. Despite this progress,
many questions still remain about the mechanisms and universal features
of ionic transport in these systems. In this paper, we examine the
use of nanopores to probe ion transport and to construct functional
nanoscale devices. Specifically, we focus on the newly predicted phenomenon
of quantized ionic conductance in nanopores as a function of the effective
pore radius - a prediction that yields a particularly transparent
way to probe the contribution of dehydration to ionic transport. We
study the role of ionic species in the formation of hydration layers
inside and outside of pores. We find that the ion type plays only
a minor role in the radial positions of the predicted steps in the
ion conductance. However, ions with higher valency form stronger hydration
shells, and thus, provide even more pronounced, and therefore, more
easily detected, drops in the ionic current. Measuring this phenomenon
directly, or from the resulting noise, with synthetic nanopores would
provide evidence of the deviation from macroscopic (continuum) dielectric
behavior due to microscopic features at the nanoscale and may shed
light on the behavior of ions in more complex biological channels. 
\end{abstract}
\maketitle

\section{Introduction}

The behavior of water and ions confined in nanoscale geometries is
of tremendous scientific interest. On the one hand, biological ion
channels, which form from membrane proteins, perform crucial functions
in the cell \cite{Hille01-1,Ashcroft00-1}. On the other hand, there
have been recent advances in aqueous nanotechnology such as nanopores
and nanochannels, which hold great promise as the basic building blocks
of molecular sensors, ultra-fast DNA sequencers, and probes of physical
processes at the nanoscale \cite{Zwolak08-1}. Indeed, nanopore-based
proposals for DNA sequencing range from measuring transverse electronic
currents driven across DNA \cite{Zwolak05-1,Lagerqvist06-1,Lagerqvist07-1,Lagerqvist07-2,Krems09-1}
to voltage fluctuations of a capacitor \cite{Heng2005-1,Gracheva2006-1,Gracheva2006-2}
to ionic currents \cite{Kasianowicz1996-1,Akeson1999-1,Deamer2000-1,Vercoutere2001-1,Deamer2002-1,Vercoutere2002-1,Vercoutere2003-1,Winters-Hilt2003-1}.

Recent experiments show that we are tantalizingly close to realizing
a device capable of ultra-fast, single-molecule DNA sequencing with
nanopores: identification of individual nucleotides using transverse
electronic transport \cite{Tsutsui10-1,Chang10-1} has been demonstrated.
Discrimination of nucleotides using their ionic blockade current when
driving them individually though a modified biological pore has also
been demonstrated \cite{Clarke09-1,Stoddart09-1}. In these systems,
the presence of water and ions will affect the signals and noise measured
and thus understanding their behavior is an important issue in both
science and technology.

Many computational studies have been dedicated to relating the three-dimensional
structure \cite{Doyle98-1,Hille01-1,Chung07-1} of biological ion
channels to their physiological function, e.g., ion selectivity. For
instance, recent studies have examined the role of ligand coordination
in potassium selective ion channels \cite{Thomas07-1,Varma07-1,Fowler08-1,Dudev09-1}.
Biological channels, however, are complex pores with many potential
factors contributing to their operation. Thus, only in a limited number
of cases have universal mechanisms of ion transport been investigated,
such as the recent work on the role of {}``topological constraints''
in ligand coordination \cite{Bostick07-1,Yu09-1,Bostick09-1}.

\begin{figure}[H]
\begin{centering}
\includegraphics[width=8cm]{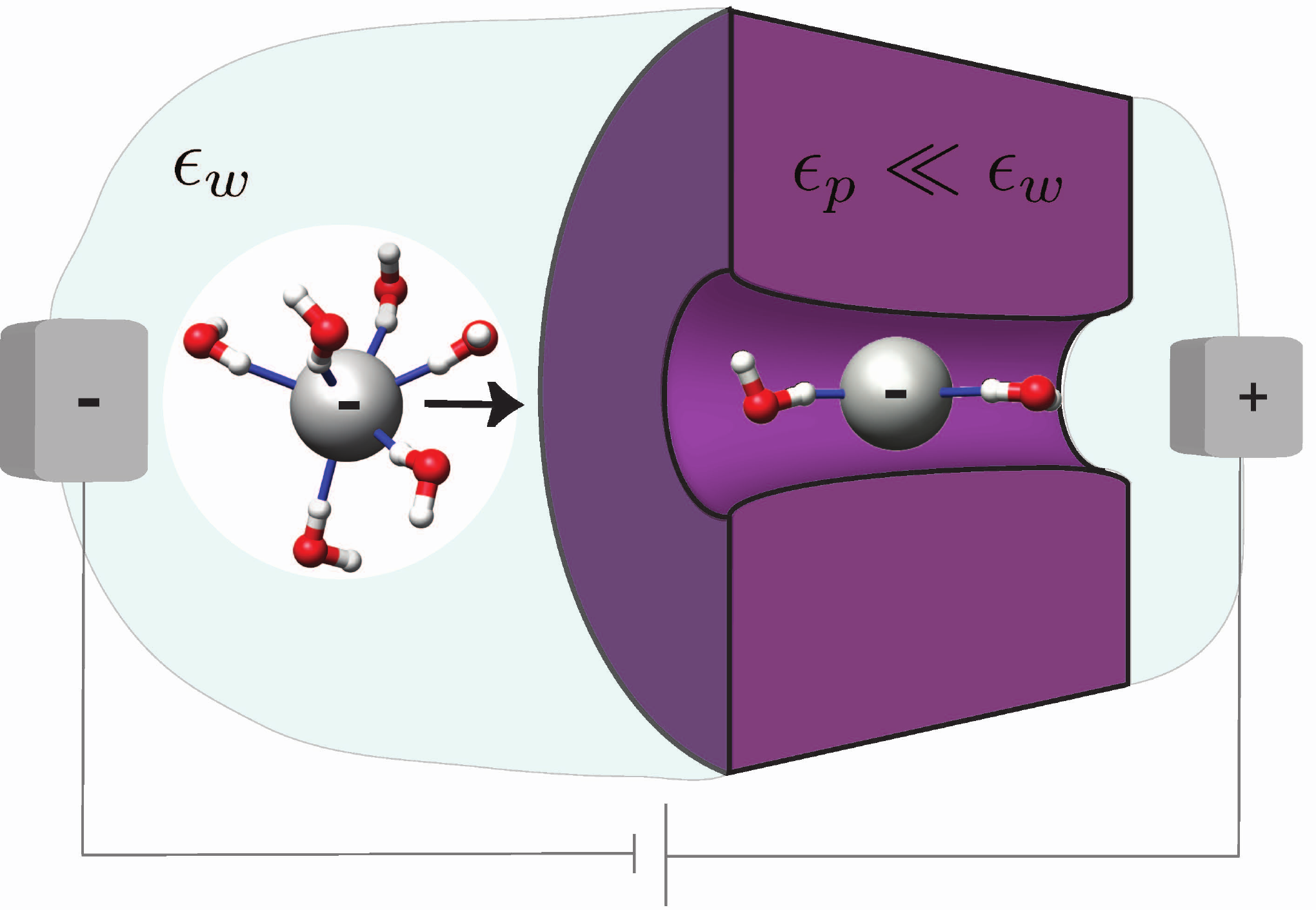} 
\par\end{centering}

\caption{Schematic of ion transport in the presence of hydration layers. Only
the first hydration layer is shown for simplicity. Ions in bulk water
form hydration layers that make the ion behave as a {}``quasi-particle''
that includes the ion and tightly bound water molecules. This quasi-particle
is then solvated in the high-dielectric water. As the ion goes from
the bulk solution to the pore it has to partially shed its hydration
layers, i.e., the quasi-particle has to break apart. This gives a
nonlinearity in the energetic barrier to transport. A continuum picture
neglects these features and considers only the dielectric barrier
that ions have to overcome by moving from bulk water with dielectric
constant $\epsilon_{w}$ into the inhomogeneous, low-dielectric pore
environment with $\epsilon_{p}\ll\epsilon_{w}$. Specialized proteins
facilitate this process in biological ion channels via the presence
of localized surface charges/dipoles and other mechanisms.\label{fig:SchematicFirst}}

\end{figure}

Fundamental developments in the fabrication of synthetic nanopores
\cite{Li2001-1,Miller01-1,Storm2003-1,Holt04-1,Siwy04-1,Harrell04-1,Holt06-1,Dekker07-1},
however, open new venues for investigating the behavior of ion channels
and dynamical phenomena of ions, (bio-)molecules, and water at the
nanoscale. For instance, what are the dominant mechanisms determining
ionic currents and selectivity? What role do binding sites play versus
hydration in constrained geometries? How accurate are {}``equilibrium''
and/or continuum theories of ion transport? Well-controlled synthetic
pores can be used in this context to examine how ion transport is
affected, for instance, by changing only the pore radius, in the absence
of binding sites and significant surface charge within the pore.

In this paper we examine the role of dehydration in ionic transport
through nanopores. In particular, we investigate the recent prediction
of quantized ionic conductance by two of the present authors (MZ and
MD) \cite{Zwolak09-1}, namely that drops in the conductance, as a
function of the effective pore radius, should occur when successive
hydration layers are prevented from entering the pore. This effect
is a classical counterpart of the electronic quantized conductance
one observes in quantum point contacts as a function of their cross
section (see, e.g., Ref.~\cite{Diventra2008-1}). We examine different
ions, both positive and negative, and of different valency (namely,
$\Cl$, $\Na$, $\K$, $\Ca$, and $\Mg$). We find that the ion type
plays only a minor role in the radii of the hydration layers, and
thus does not affect much the pore radii at which a sudden drop in
the current is expected. Divalent ions, however, are the most ideal
experimental candidates for observing quantized ionic conductance
because of their more strongly bound hydration layers. Further, the
fluctuating hydration layer structure and changing contents of the
pore should give a peak (versus the effective pore radius) in the
relative current noise - giving an additional method to observe the
effect of the hydration layers. Thus, we elucidate how quantized ionic
conductance provides a novel tool to deconstruct the energetic contributions
to ion transport.

The paper is organized as follows: In Sec. \ref{sec:Macro}, we give
a macroscopic (i.e., a continuum electrostatic) viewpoint on the energetics
of ion transport. In Sec. \ref{sec:Hydration}, we examine how ions
induce local structures in the surrounding water known as hydration
layers - an effect that is not taken into account when using continuum
electrostatics to estimate energetic barriers to transport. Further,
we calculate the energies stored in these layers and develop a model
for the energetic barrier for ions entering a pore. In Sec. \ref{sec:Ionic-Currents},
we use a Nernst-Planck approach to relate this barrier to the ionic
current. In Sec. \ref{sec:Noise}, we discuss how the presence of
the hydration layers gives rise to a peak in the relative noise in
the ionic current at values of the effective pore radius congruent
with a layer radius. In Sec. \ref{sec:Conclusions}, we then present
our conclusions.

\section{Ionic Transport\label{sec:Macro}}

The experimental set-up we are interested in is that of ions driven
through a pore/channel of nanoscale dimensions under the action of
a static electric field %
\footnote{Although similar conclusions should apply in other scenarios such
as the generation of a concentration gradient.%
}. Such a situation is depicted in Fig. \ref{fig:SchematicFirst}.
A simple approach to ionic transport is to envision the ions moving
through an energetic barrier due to going from the high-dielectric
aqueous environment into the inhomogeneous, low-dielectric environment
of the pore, treating the surroundings as continuum media. The resulting
approach is inherently static: by analyzing the energetic barrier
to (near-equilibrium) transport one obtains information about how
different factors - the pore material (through its dielectric constant),
the pore dimensions, the presence of surface charges, and the presence
of the high-dielectric water along the pore axis - would affect transport.

Indeed, one of the first calculations of the dielectric barrier (using
a {}``Born solvation'' model) was done by considering the ion solvated
in water and moved into a low-dielectric, pore-less membrane \cite{Parsegian69-1,Parsegian75-1}.
This provides an estimate of the energies involved by calculating
the energy change of solvating the ion in continuum water, with dielectric
constant $\epsilon_{w}\approx80$, to {}``solvating'' it in a continuum
material with $\epsilon_{p}\approx2$ (representative of lipid membranes
\footnote{In pores - especially biological pores - the membrane dielectric constant,
$\epsilon_{m}$, and the dielectric constant of the pore material,
$\epsilon_{p}$, can be different.%
}). For instance, the energy change of a $\Cl$ ion, with effective
radius $R\approx2\,\textrm{\AA}$ %
\footnote{The effective radius can be estimated from, e.g., molecular dynamics
simulations that give a surface where the screening charge due to
the hydrogen or oxygen atoms of water fluctuates. For instance, Figs.
\ref{fig:monoden} and \ref{fig:diE} show this surface (see also
Refs. \cite{Rashin85-1,Roux90-1}). %
}, moved from continuum water to the continuum material is \begin{eqnarray}
\Delta U & = & \frac{e^{2}}{8\pi R\epsilon_{0}}\left(\frac{1}{\epsilon_{p}}-\frac{1}{\epsilon_{w}}\right)\label{eq:BornEst}\\
 & \approx & 1.8\eV.\label{eq:BornEstSpe}\end{eqnarray}
 This is quite a substantial energy change - about half the solvation
free energy of $\Cl$ \cite{Hille01-1,Marcus91-1}. The finite thickness
of the membrane does not change this value significantly. For thick
membranes, it is lowered by \cite{Parsegian69-1,Parsegian75-1} \[
\frac{e^{2}}{4\pi\epsilon_{0}\epsilon_{p}l}\ln\left(\frac{2\epsilon_{w}}{\epsilon_{w}+\epsilon_{p}}\right),\]
 for $\epsilon_{w}\gg\epsilon_{p}$, where $l$ is the membrane thickness
(and pore length). For $\epsilon_{p}\approx2$ and $\epsilon_{w}\approx80$,
this gives $\sim5/l\eV\,\textrm{\AA}\approx0.1\eV$ for a membrane
of thickness $l=50\,\textrm{\AA}$. That is, the Born estimate in
Eq. \eqref{eq:BornEst} is lowered to $\sim1.7\eV$. However, the
membrane width \cite{Levitt78-1} and composition can play a significant
role in this estimate. For the common synthetic pores made of silicon
dioxide ($\epsilon_{p}\approx4$) or silicon nitride ($\epsilon_{p}\approx7.5$)
the estimate in Eq. \eqref{eq:BornEst} is reduced from $\sim1.8\eV$
to $\sim0.9\eV$ and $\sim0.4\eV$, respectively. These barriers are
more than an order of magnitude larger than $k_{B}T$ at room temperature,
where $k_{B}$ is the Boltzmann constant.

Due to this magnitude, it is clear that the energy scale of solvation
is one of the controlling factors in ion transport. However, in addition
to the above there is water present in the pore. One expects, therefore,
that the energy of solvation would be decreased from simple estimates
like that of Eq. \eqref{eq:BornEst}. Several groups have calculated
this contribution \cite{Parsegian69-1,Parsegian75-1,Levitt78-1,Jordan81-1,Jordan82-1}.
For instance, Ref. \cite{Roux04-1} shows that the energy barrier
of bringing an ion from continuum water into a low-dielectric, continuum
membrane is reduced from $\sim40\,\mathrm{kcal/mol}\approx1.7\,\mathrm{eV}$
to $\sim20\,\mathrm{kcal/mol}\approx0.9\,\mathrm{eV}$ by the presence
of water in the pore. This demonstrates that a pore filled with a
high dielectric medium (e.g., continuum water) can significantly lower
the barrier to transport. Even still, the barrier remains substantial.

In biological systems, however, the pores provide a channel with a
much lower barrier as indicated by the conductance of many biological
ion channels. These pores are formed from specialized proteins whose
role is precisely to facilitate passage of ions (and further to selectively
allow passage of certain ions). Clearly, pores with internal charges
and/or dipoles can significantly reduce the energetic barriers for
transport. Indeed, the effect of surface charges has been calculated
in clean pores \cite{Teber2005-1,Zhang2005-1,Kamenev2006-1} and when
present in sufficient amounts would negate the effect we predict as
the reduction of the energetic barrier would be comparable to, or
larger than, the hydration layer energies. Therefore, our interest
is in clean pores with little to no surface charge where clear-cut
experiments can be performed to understand the effect of hydration
on transport. This rules out the direct use of some biological ion
channels, particularly those with very small pores where single-file
transport occurs \cite{Finkelstein81-1,Doyle98-1,Morais01-1}, because
of the presence of localized charges and dipoles.

To conclude this section, we note that the continuum description suffers
from a number of issues at the nanoscale: it is only valid beyond
the correlation length of the material \cite{JacksonWater}, which
for the strong fields around an ion is $\sim8\,\textrm{\AA}$ for
water (see below), similar to the $\sim5-8\,\textrm{\AA}$ in water
only \cite{Narten72-1}; linear continuum electrostatics is only valid
when the polarization field is co-linear with the electric field (not
the case in the hydration layers we discuss below); in a related issue,
it is only valid for weak fields (in the context of ion channels,
see, for example, Sec. 3.4 in Ref. \cite{Roux04-1}); there is also
an issue of where the {}``surface'' separating the charge and the
dielectric membrane/continuum water is located, especially for fluctuating
atomic ensembles as is the case for protein pores and molecular (rather
than continuum) water. Thus, while a continuum picture can highlight
some general features of the energetic barrier to ion transport -
in some cases giving compact analytical expressions - it breaks down
when trying to understand the effect of structure at the nanoscale.
In fact, macroscopic, continuum electrostatics is not designed to
study specific features or short-range interactions at these length
scales. This is precisely what we seek to address in the following
sections.

\section{Hydration of Ions\label{sec:Hydration}}

We begin our study of quantized conductance by first illustrating
how ions are hydrated in solution and then discuss the energies involved
in this process. The formation of hydration layers around ions has
been known for some time (see, e.g., Ref. \cite{Hille01-1}), and
is due to the strong local electric field around the ion and to repulsive
short-range interactions among molecular/atomic species. We use molecular
dynamics (NAMD2 \cite{Phillips05-1}) simulations to understand the
structure of hydration layers when different ions are inside and outside
of nanopores %
\footnote{For an ion in bulk water, we simulated a hexagonal box of $150\,\textrm{\AA}$
height and $43\,\textrm{\AA}$ radius with periodic boundary conditions
in all directions. We then fixed an ion in the center of the box and
counterion(s) near the edge of the box, far away from the ion of interest.
For an ion in a pore, a cylindrical pore of radius $R$ was cut into
a hexagonal silicon nitride film $97\,\textrm{\AA}$ thick and of
$29\textrm{\AA}$ radius. This was accomplished by removing all silicon
and nitrogen atoms within a distance $R$ from the z-axis. An ion
was fixed in the middle of the pore and counterion(s) were fixed outside
of the pore. The system was then solvated in water resulting in a
box of linear dimension $167\,\textrm{\AA}$ in the z-direction. An
energy minimization procedure was then run, the system was heated
up to 295 K, and finally the production run started. The first 600
ps were discarded to remove artifacts from the initial conditions
and the information from the subsequent 2 ns collected. Other simulation
details are as in Ref.~\cite{Krems09-1}.%
}.

\begin{figure*}[t]
\begin{centering}
\includegraphics[width=15cm]{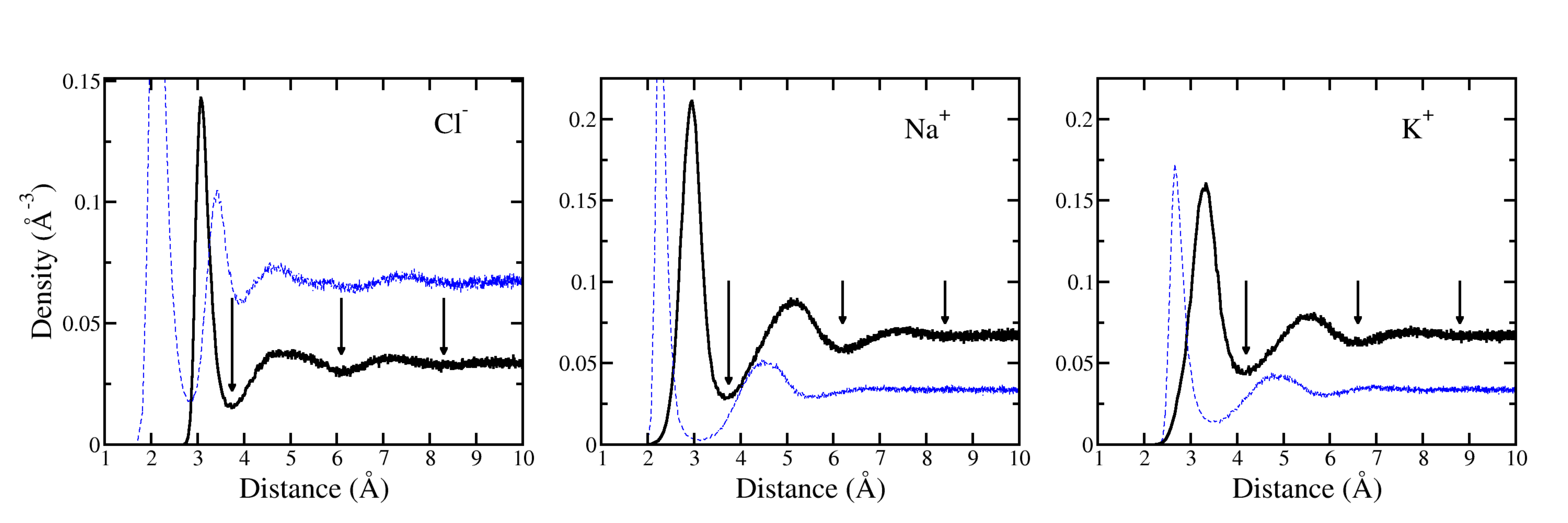} 
\par\end{centering}

\begin{centering}
\includegraphics[width=15cm]{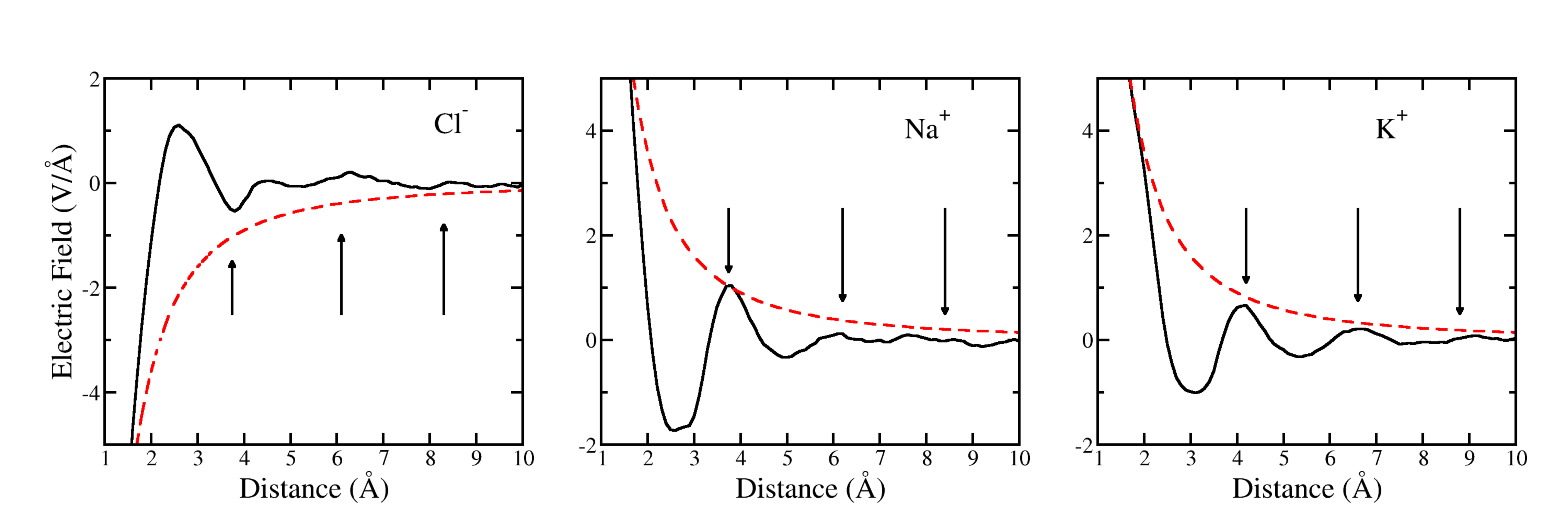} 
\par\end{centering}

\caption{Top panels: Water density oscillations versus distance for $\Cl$,
$\Na$, and $\K$ in bulk water. Black, solid lines indicate the density
calculated from the oxygen atom positions for $\Cl$ and hydrogen
atom positions for the cations. The arrows indicate the minimum in
the density oscillations. The blue, dashed lines indicate the density
calculated from the hydrogen atom positions for $\Cl$ and oxygen
atom positions for the cations. \label{fig:monoden} Bottom panels:
The electric field due to both the bare ion (red, dashed line) and
due to the ion plus partial charges on the water molecules (black,
solid). The arrows again indicate the minimum in the density oscillations.
\label{fig:monoE}}

\end{figure*}

\begin{figure*}[t]
\begin{centering}
\includegraphics[width=10cm]{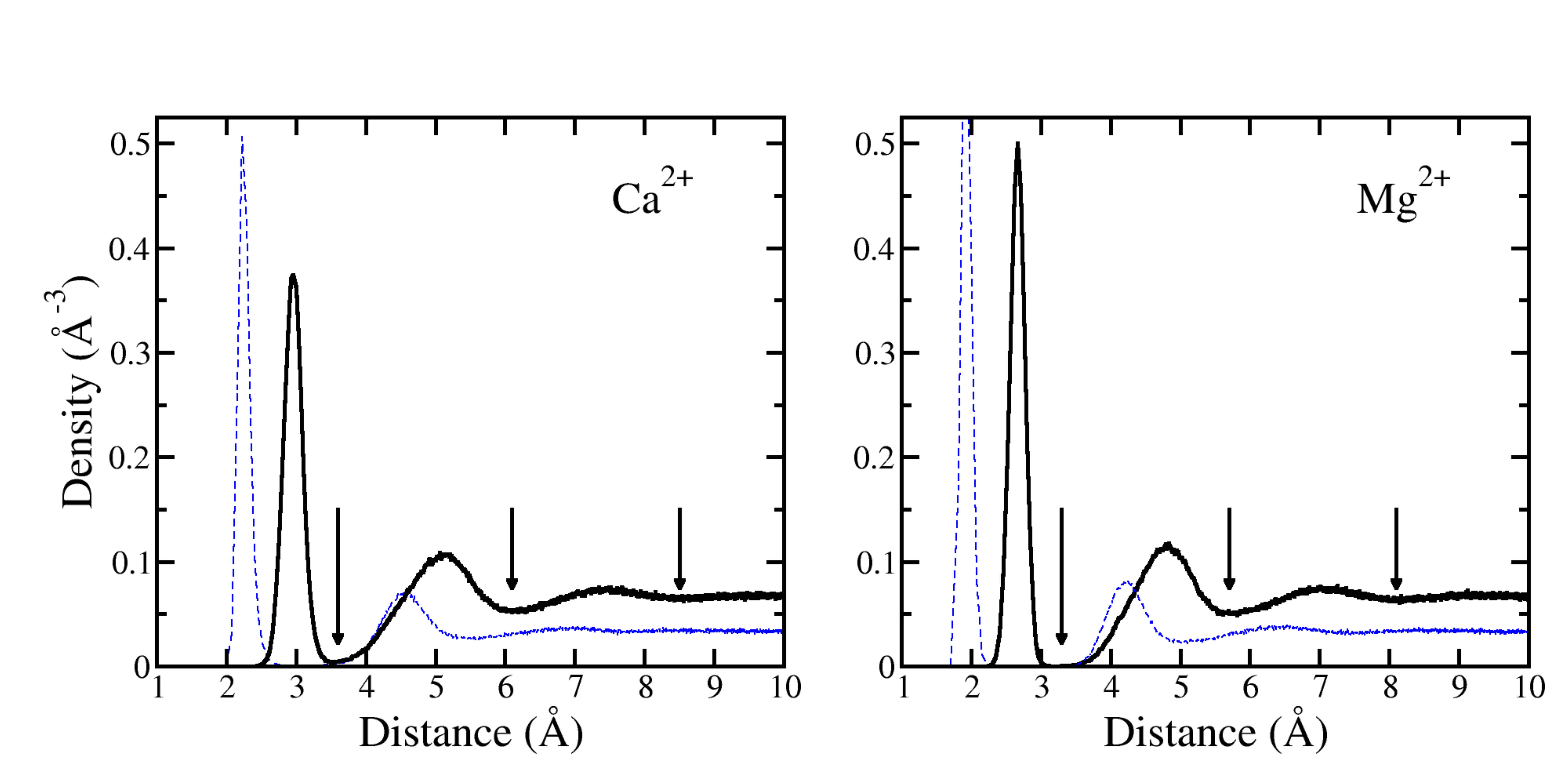} 
\par\end{centering}

\begin{centering}
\includegraphics[width=10cm]{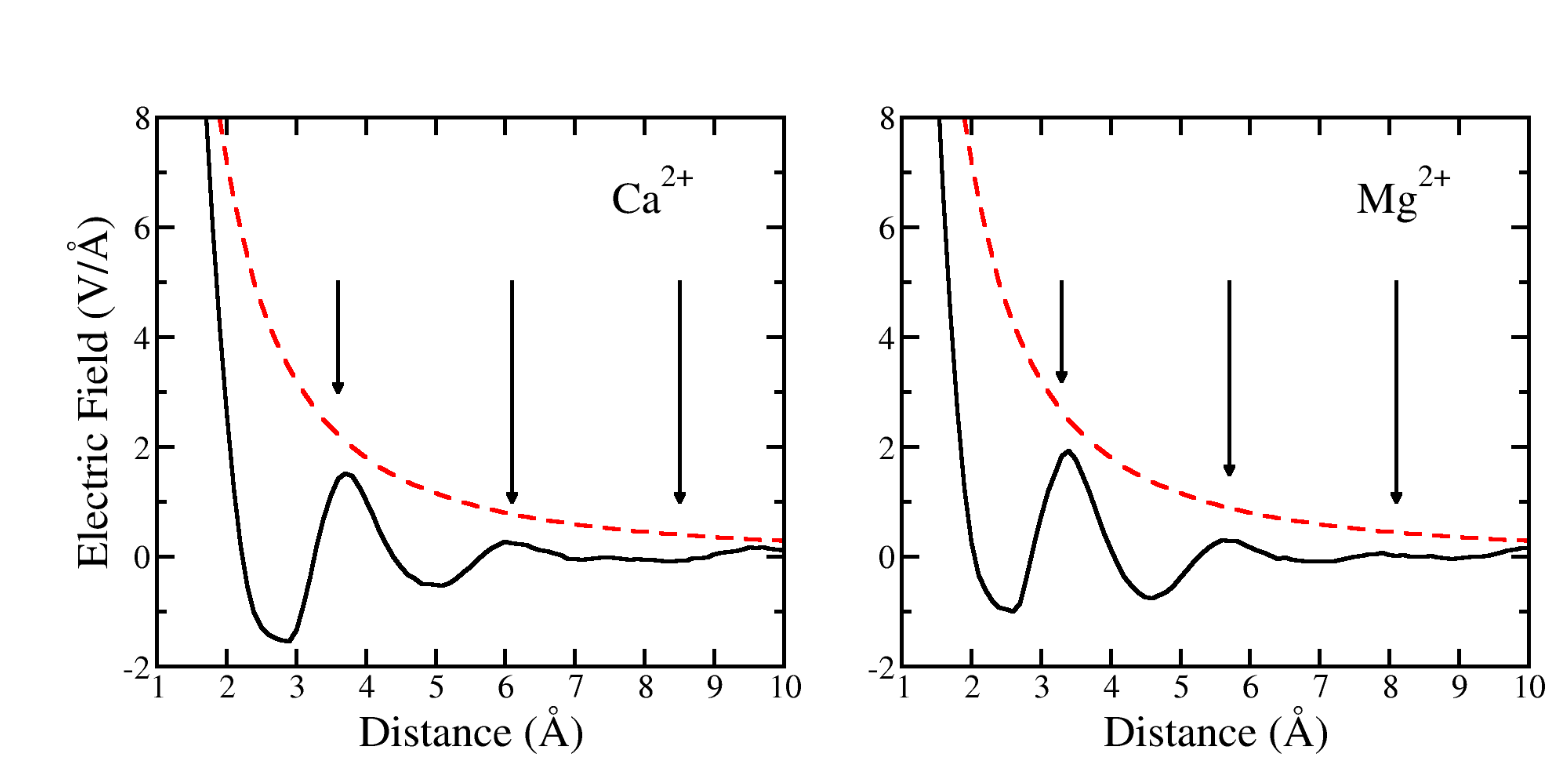} 
\par\end{centering}

\caption{Top panels: Water density oscillations versus distance for $\Ca$
and $\Mg$ in bulk water. Black, solid lines indicate the density
calculated from the hydrogen atom positions. The arrows indicate the
minimum in the density oscillations. The blue, dashed lines indicate
the density calculated from the oxygen atom positions. \label{fig:diden}
Bottom panels: The electric field due to both the bare ion (red, dashed
line) and due to the ion plus partial charges on the water molecules
(black, solid). The arrows again indicate the minimum in the density
oscillations. \label{fig:diE}}

\end{figure*}

Figures \ref{fig:monoden} and \ref{fig:diden} show the water density
oscillations for several common ionic species %
\footnote{We calculated the density of water surrounding each ion by placing
$1\,\textrm{\AA}^{3}$ shells concentric with the ion. The inner shells
have a larger width to give the same volume. We then counted the number
of atoms (either hydrogen or oxygen) within each of the shells throughout
the 2 ns simulation at time intervals of 200 fs. Due to the smaller
bin sizes, the plots have minor differences from Ref. \cite{Zwolak09-1}
at small distances from the ion.%
}. There is a strong peak in water density about $3\,\textrm{\AA}$
away from the ions, with two further oscillations after that spaced
about $2\,\textrm{\AA}$ apart. These oscillations signify that there
are strongly bound water molecules forming around the ions. Table
\ref{tab:values} lists the hydration layer radii from both this study
and experiment. We find very good agreement with the experimental
data for all cases. The water density approaches the bulk value ($\sim0.033\,\mathrm{Molecules}/\textrm{\AA}^{3}$)
at about $10\,\textrm{\AA}$, which is also consistent with the experimental
value.

\begin{table*}[t]
\begin{centering}
\begin{tabular}{c|c|c|c|c|c|c|c}
Ion  & $R_{i}$ ($\textrm{\AA}$) (th)  & $R_{XO}$ ($\textrm{\AA}$) (th)  & $R_{XO}$ ($\textrm{\AA}$) (exp)  & $R_{I},R_{O}$ ($\textrm{\AA}$)  & $-U_{i}$ ($\mathrm{eV}$) (th)  & $-\Delta G$ ($\mathrm{eV}$) (exp)  & $\mu$ ($\mathrm{m}^{2}V/s$) (exp) \tabularnewline
\hline
\hline 
$\Cl$  & 3.1, 4.9, 7.1  & 3.1  & 3.19  & 2.0, 3.9, 6.2, 8.5  & 1.73, 0.68, 0.31  & 3.54  & $7.92\times10^{-8}$\tabularnewline
\hline 
$\Na$  & 2.9, 5.1, 7.5  & 2.3  & 2.44  & 1.9, 3.8, 6.2, 8.4  & 1.51, 0.72, 0.30  & 3.80  & $5.19\times10^{-8}$\tabularnewline
\hline 
$\K$  & 3.3, 5.6, 7.8  & 2.7  & 2.81  & 2.4, 4.2, 6.6, 8.8  & 1.15, 0.61, 0.27  & 3.07  & $7.62\times10^{-8}$\tabularnewline
\hline 
$\Ca$  & 3.0, 5.1, 7.5  & 2.2, 4.6  & 2.42, 4.55  & 1.8, 3.6, 6.1, 8.5  & 7.89, 3.23, 1.32  & 15.65  & $6.17\times10^{-8}$\tabularnewline
\hline 
$\Mg$  & 2.7, 4.8, 7.1  & 1.9, 4.2  & 2.09, 4.20  & 1.5, 3.3, 5.7, 8.1  & 10.33, 3.62, 1.48  & 19.03  & $5.5\times10^{-8}$\tabularnewline
\hline
\end{tabular}
\par\end{centering}

\caption{Table of physical quantities from simulation and experiment. The theoretical
hydration layer radii, $R_{i}$, for all three layers are defined
using the ion-oxygen distance for $\Cl$ and ion-hydrogen distance
for the cations. The first oxygen density maximum, $R_{XO}$, is for
all ions $X$ using the present theory and experiment (average values
taken from Ref. \cite{Ohtaki93-1}). The second maximum is also shown
for the divalent ions from both theory and experiment. The inner/outer
radii that enter Eq. \eqref{eq:LayerU} are shown, the first of which
is calculated such that Eq. \eqref{eq:LayerU} equals $\Delta G$
(exp) when $R_{i\nu}^{O}\rightarrow\infty$ (see also text). The next
three inner/outer radii are taken from the minima of the oxygen density
for $\Cl$ and the minima of the hydrogen density for the cations.
Further, we report the layer energies $U_{i}$ (using $\epsilon_{p}=1$),
and the Gibb's free energy from experiment \cite{Marcus91-1}, and
the experimental mobilities \cite{Hille01-1} used in this work. \label{tab:values}}

\end{table*}

The oscillations in water density also give rise to oscillations in
the local electric field. Figures \ref{fig:monoE} and \ref{fig:diE}
show this for monovalent and divalent ions where the time-averaged
electric field was calculated from the bare ion value plus a sum over
all partial charges given by the hydrogen and oxygen atoms of water
\footnote{The electric field was calculated by summing the contributions from
the ion and all partial charges (on the hydrogen and oxygen of the
water) within $15\,\textrm{\AA}$ from every field point. The angular
component to the field was several orders of magnitude smaller because
the time-averaged field has essentially spherical symmetry. In Ref.
\cite{Zwolak09-1} all water molecules were modeled as dipoles.%
}. In the figures, the first hydration layer gives pronounced field
oscillations for all species examined. The other oscillations in the
field are more well-defined for $\K$, $\Ca$, $\Mg$, and to some
extent $\Na$, compared to $\Cl$. Anions, such as $\Cl$, have a
different structure of the water around them compared to cations:
in the first layer, they pull one of the hydrogen atoms of each of
the water molecules closer while the other interferes with the formation
of the second layer, possibly hindering the ability of the second
layer to form a {}``perfect'' screening surface. The fact that the
electric field is not simply suppressed by $1/\epsilon_{w}$ shows
the difficulty of a macroscopic (continuum) dielectric picture to
predict behavior at the nanoscale (similar to well-known features
in other systems such as Friedel oscillations and apparent from the
derivation of continuum electrostatics, where averaging is required
over length scales much larger than the correlation length of the
material \cite{JacksonWater}).

\begin{figure}
\begin{centering}
\includegraphics[width=8cm]{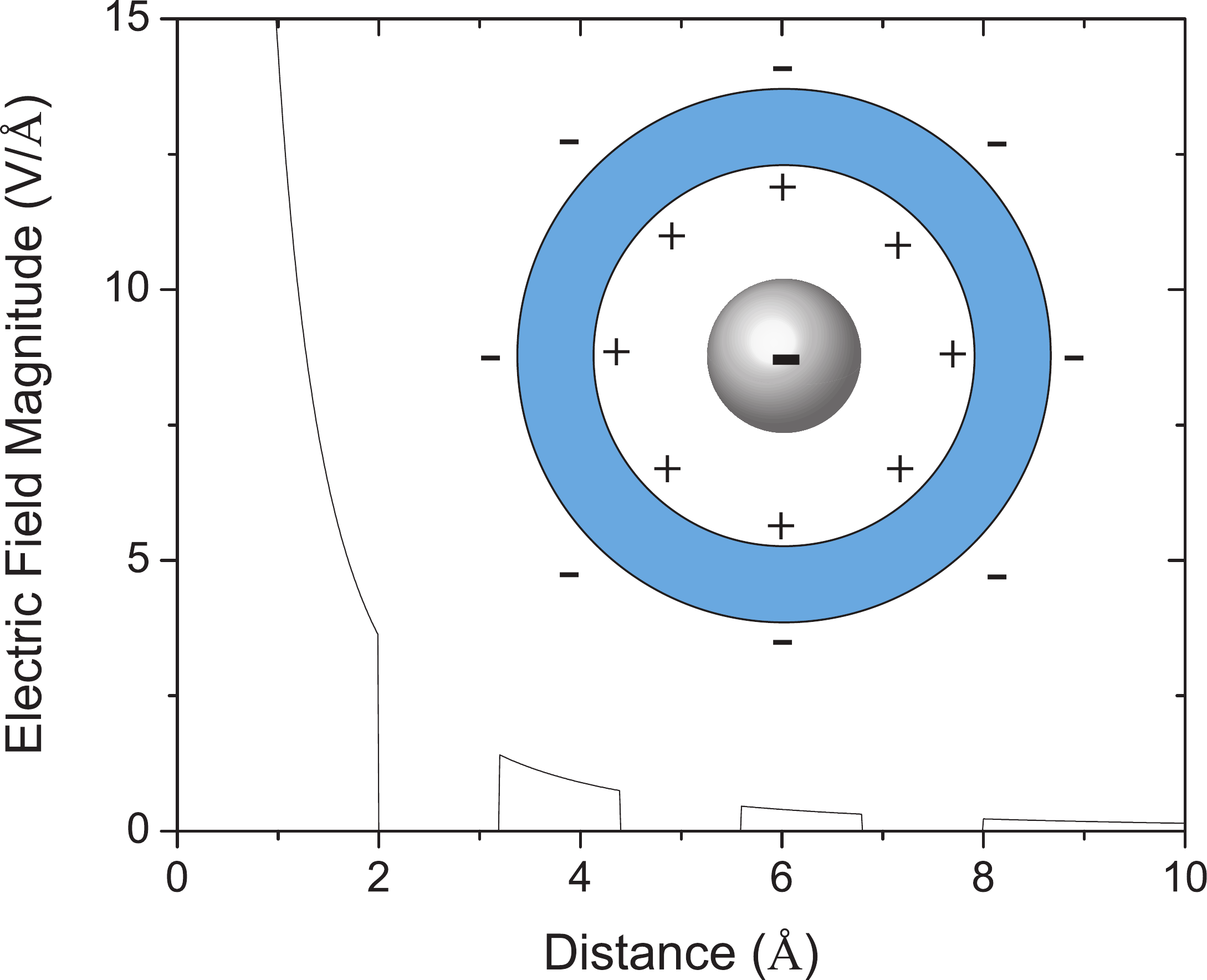} 
\par\end{centering}

\caption{The magnitude of the approximate electric field (given by a set of
Gauss surfaces, see inset) used to represent the fields in Figs. \ref{fig:monoE}
and \ref{fig:diE}, and also used to calculate the energy contained
in each layer (Eq. \eqref{eq:LayerU}). \label{fig:Gauss}}

\end{figure}

We now estimate the energies contained in these layers, which we list
in Table \ref{tab:values}. The electric fields seen in Figs. \ref{fig:monoE}
and \ref{fig:diE} show an oscillating behavior that is reminiscent
of a set of Gauss surfaces, i.e., layers of alternating charge that
screen the field of the ion. Thus, in order to estimate the energies
contained in the layers, we replace the microscopic structure giving
rise to the complex field by a set of surfaces as shown in Fig. \ref{fig:Gauss}
that perfectly screen (with dielectric constant $\epsilon_{w}$),
rather than over-screen, the ion charge.

Within this picture, the energy of the $i^{th}$ hydration layer of
ionic species $\nu$ is \cite{Zwolak09-1} \begin{equation}
U_{i\nu}^{o}=\frac{q_{\nu}^{2}}{8\pi\epsilon_{0}}\left(\frac{1}{\epsilon_{p}}-\frac{1}{\epsilon_{w}}\right)\left(\frac{1}{R_{i\nu}^{O}}-\frac{1}{R_{i\nu}^{I}}\right),\label{eq:LayerU}\end{equation}
 where $q_{\nu}$ is the ionic charge and $R_{i\nu}^{I\,(O)}$ are
the inner (outer) radii demarcating the hydration layer as obtained
from the water density oscillations. In order to obtain the innermost
radius we compute the total solvation energy, $U_{T}=-q_{\nu}^{2}\left(\epsilon_{w}-\epsilon_{p}\right)/\left(8\pi\epsilon_{0}\epsilon_{p}\epsilon_{w}R_{1\nu}^{I}\right)$,
and compare with the experimental free energies \cite{Marcus91-1},
which are dominated by the electrostatic energy. These free energies,
together with the layer energies (for $\epsilon_{p}=1$), are tabulated
in Table \ref{tab:values}. Except for the third hydration layer for
monovalent ions, the layer energies are greater than other free energy
contributions such as the entropy change due to the water structure
or van der Waals interactions \cite{Yang07-1,Ignaczak99-1}. In Eq.
\eqref{eq:LayerU} we have also added a possible screening contribution,
$\epsilon_{p}$, from the pore material and/or charges on the surface
of the pore. In Ref. \cite{Zwolak09-1} this was assumed to be one:
the low-dielectric pores reduce the energy barrier only by a small
amount and in a different functional form. In Sec. \ref{sec:Noise}
we will discuss the effect of this screening on the detection of quantization
steps.

Previously, we proposed a model for how the energy is depleted in
a hydration layer as the effective radius of the pore, $R_{p}$, is
reduced \cite{Zwolak09-1}. In this model, the energy change is proportional
to the remaining surface area of a hydration layer within a pore.
It takes into account both that the water-ion interaction energy of
small water clusters is approximately linear in the number of waters
\cite{Ignaczak99-1,Kistenmacher74-1} and that molecular dynamics
simulations show a time-averaged water density with partial spherical
shells when an ion is inside a pore of small enough radius (see Fig.
1 in Ref. \cite{Zwolak09-1}). Contributions from, e.g., van der Waals
interactions with the pore and changes in the water-water interaction,
are small \cite{Yang07-1,Ignaczak99-1}. Thus, the energy of the remaining
fraction $f_{i\nu}$ of the $i^{th}$ layer in the pore is taken as
$U_{i\nu}=f_{i\nu}U_{i\nu}^{o}$. The fraction of the layer intact
is $f_{i\nu}=S_{i\nu}/4\pi R_{i\nu}^{2}$ where $S_{i\nu}$ is the
surface area (of the spherical layer) remaining where the water dipoles
can fluctuate. The latter is given by \begin{equation}
S_{i\nu}=2\Theta\left(R_{i\nu}-R_{p}\right)\int_{0}^{2\pi}d\phi\int_{0}^{\theta_{c\nu}}d\theta R_{i\nu}^{2}\sin\theta,\end{equation}
 where $\Theta\left(x\right)$ is the step function and $\theta_{c\nu}=\sin^{-1}(R_{p}/R_{i\nu})$.
When $R_{p}<R_{i\nu}$, the fraction of the surface left is \begin{equation}
f_{i\nu}\left(R_{p}\right)=1-\sqrt{1-\left(\frac{R_{p}}{R_{i\nu}}\right)^{2}}.\end{equation}
 The total internal energy change will then result from summing this
fractional contribution over the layers to get \begin{equation}
\Delta U_{\nu}(R_{p})=\sum_{i}(f_{i\nu}(R_{p})-1)U_{i\nu}^{o}.\label{eq:ElectroBarrier}\end{equation}

We stress first that the effective radius $R_{p}$ is not necessarily
the nominal radius defined by the pore atoms. Rather, it is the one
that forces the hydration layer to be partially broken because it
can not fit within the pore, and it could be smaller than the nominal
pore radius by the presence of, e.g., a layer of tightly bound water
molecules on the interior surface of the pore. Second, our model misses
internal features of the hydration layers themselves. For instance,
Ref. \cite{Song09-1} examines the first hydration layer structure
in carbon nanotubes of different radii. These authors find a large
increase in the energy barrier when the pore radius nears the inner
hydration layer. They also seem to observe \emph{sub-steps} in the
water coordination number within the inner shell as the pore radius
is reduced. Thus, although our model contains only a single {}``smoother''
step, experiments could very well observe these internal sub-steps
corresponding to the sudden loss of a single or few water molecules
out of a given hydration layer.

\begin{figure}
\begin{centering}
\includegraphics[width=8cm]{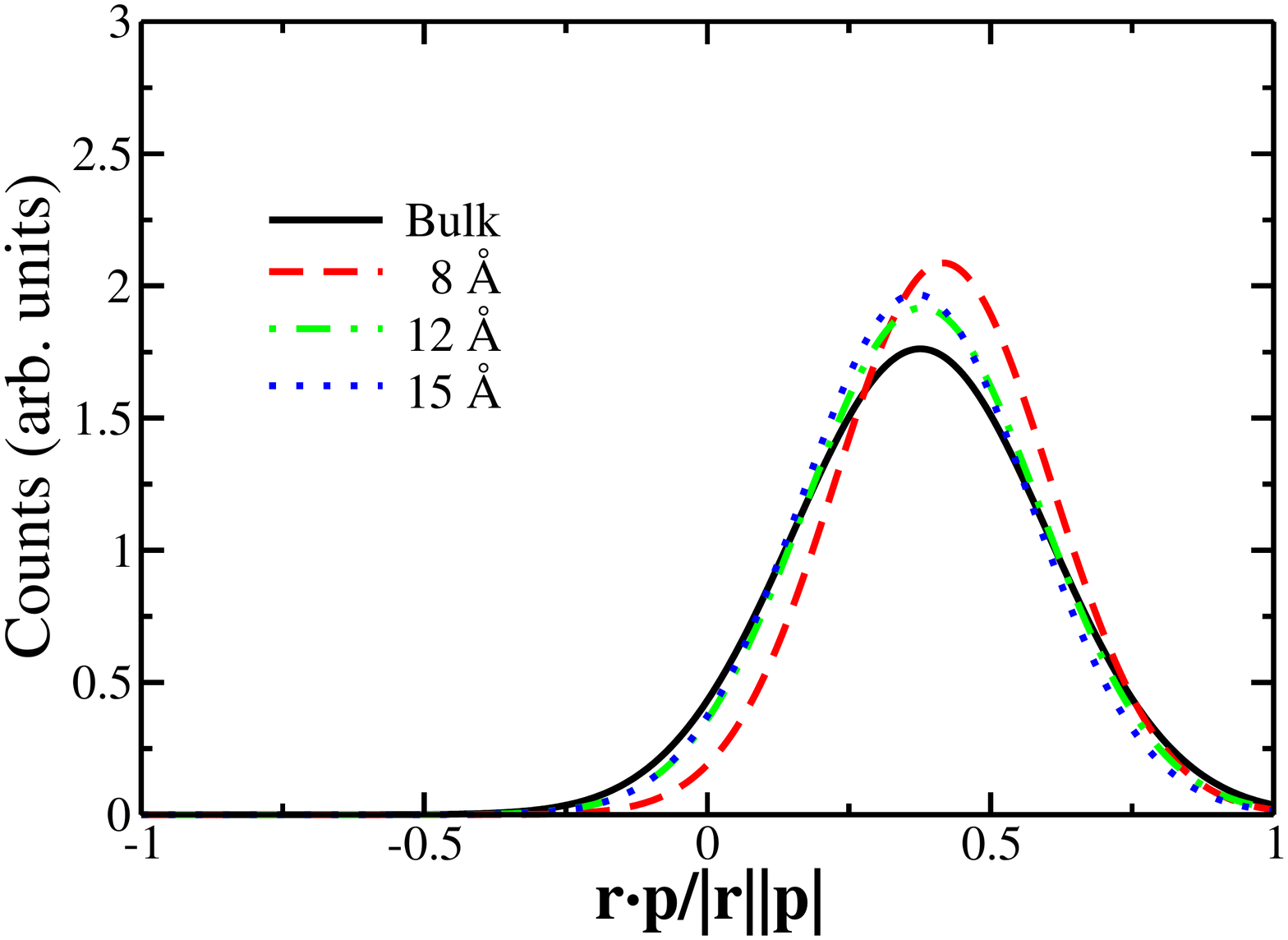} 
\par\end{centering}

\caption{Distributions of the dipole orientations of water molecules within
a cylindrical annulus $5\,\textrm{\AA}$ away from, i.e., in the second
hydration layer of, a $\Ca$ ion in bulk (black line) and inside pores
of radius $8\,\textrm{\AA}$ (red dashed line), $12\,\textrm{\AA}$
(green dash-dotted line) and $15\,\textrm{\AA}$ (blue dotted line).
The mean value is around $0.38$ (corresponding to the water dipole
pointing 68 degrees away from the ion-water vector), except for the
$8\,\textrm{\AA}$ pore, which increases to 0.42 (corresponding to
the water dipole pointing 65 degrees away from the ion-water vector).
This signifies a moderate tightening of the water dipole around the
field of the ion as the pore size is reduced. \label{fig:DipoleHisto}}

\end{figure}

Another basic assumption in our model is that the interaction energy
of the water molecules in each layer is the same regardless of whether
the ion is inside or outside of the pore. Figure \ref{fig:DipoleHisto}
shows the distribution of the dipole orientation of water molecules
both in bulk and inside pores of different radius %
\footnote{For each time step, all water molecules within a cylindrical annulus
coaxial with the $+z$-axis were taken, where the annulus has a $1.5\,\textrm{\AA}$
width, $1.5\,\textrm{\AA}$ height, and a central ring $5\,\textrm{\AA}$
from the ion. Then the unit vector connecting the oxygen atom of those
molecules to the midpoint between the hydrogen atoms (the unit dipole
$\hat{p}(t)$) was generated together with the unit position vector
of the water molecules at the centroid of the molecule $\hat{r}(t)$.
We then took the scalar product $\hat{p}\cdot\hat{r}$ per molecule,
and averaged over the molecules. This set of data was then made into
a histogram of 501 bins evenly spaced from -1 to 1.%
}. The average dipole orientation of the waters changes very little
inside the pore, as do their fluctuations, thus supporting this assumption.
In addition, the structure of the first two hydration layers (not
shown) is essentially the same in and out of the smallest ($8\,\textrm{\AA}$)
pore.

\begin{figure*}[t]
\begin{centering}
\includegraphics[width=15cm]{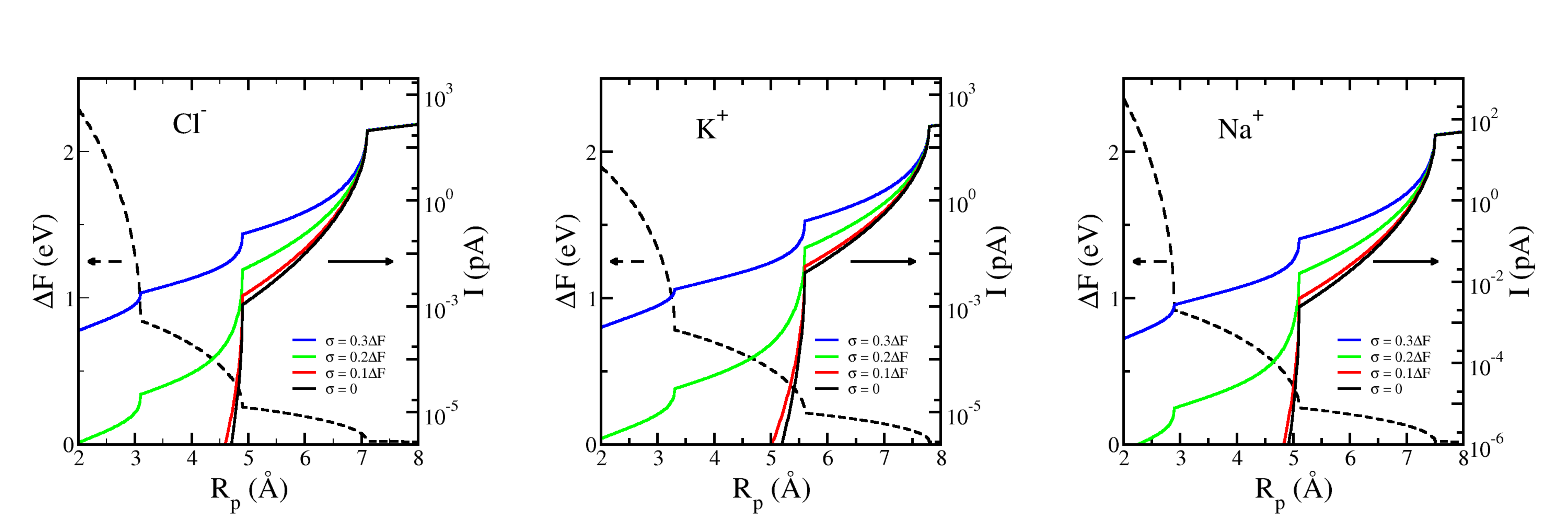} 
\par\end{centering}

\caption{Free energy changes, $\Delta F_{\nu}$, and currents versus the effective
pore radius for the monovalent ions and a field of $1\,\mathrm{mV/\textrm{\AA}}$.
The black, dashed line indicates the free energy change. The remaining
lines indicate the current with different standard deviations of the
noise (see text for details). The currents are for $\sigma=p\Delta F$,
with $p=0.3,\,0.2,\,0.1,\,0$ from top to bottom. \label{fig:EandImono}}

\end{figure*}

\begin{figure*}[t]
\begin{centering}
\includegraphics[width=10cm]{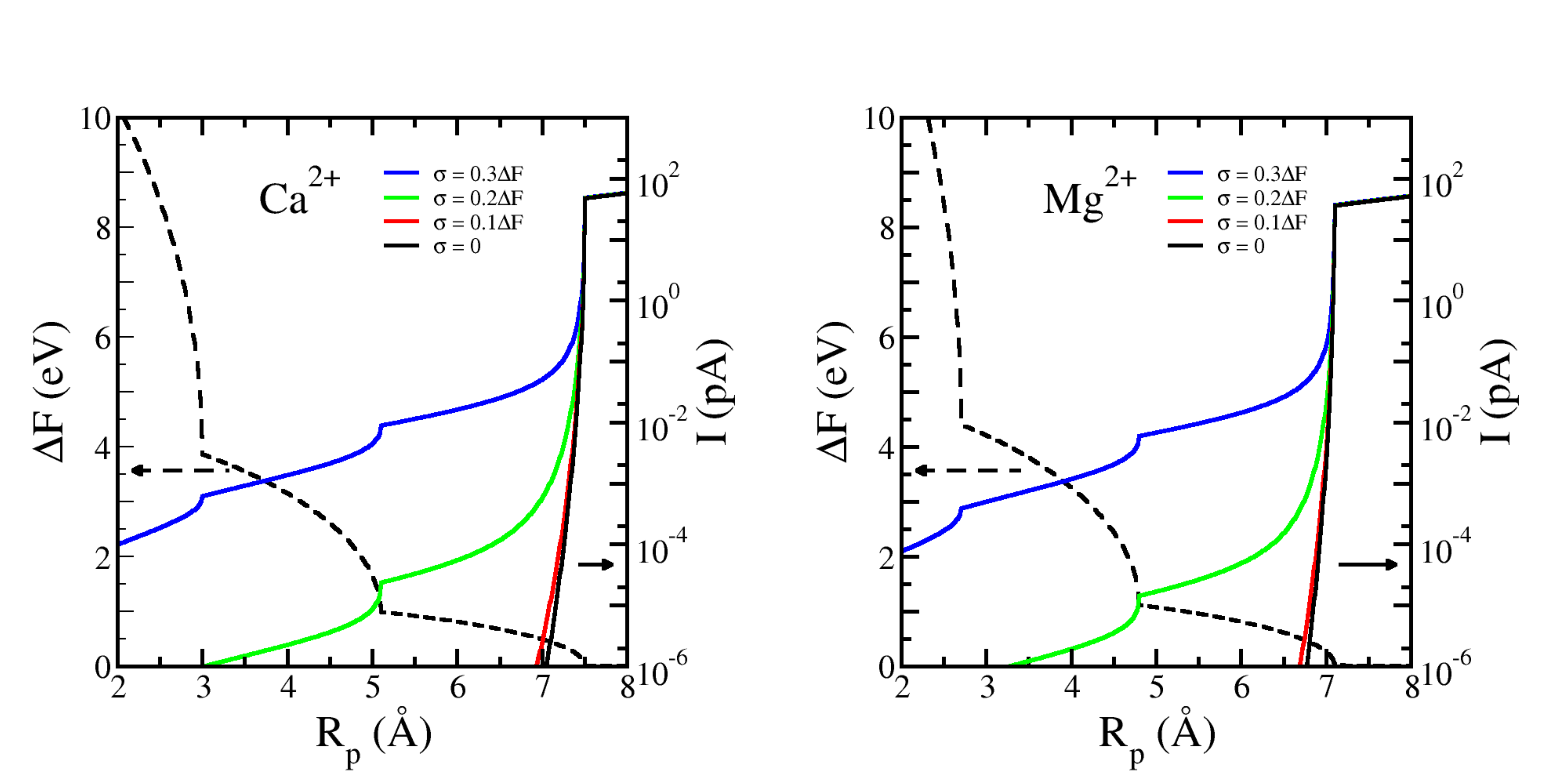} 
\par\end{centering}

\caption{Free energy changes, $\Delta F_{\nu}$, and currents versus the effective
pore radius for the divalent ions and a field of $1\,\mathrm{mV/\textrm{\AA}}$.
The black, dashed line indicates the free energy change. The remaining
lines indicate the current with different standard deviations of the
noise (see text for details). The currents are for $\sigma=p\Delta F$,
with $p=0.3,\,0.2,\,0.1,\,0$ from top to bottom. \label{fig:EandIdi}}

\end{figure*}

In order to make a connection with the ionic current (in Sec. \ref{sec:Ionic-Currents}
below), we calculate the free energy %
\footnote{Here we deal with constant volume and temperature and thus use the
Helmholtz free energy.%
} change for species $\nu$ as \begin{equation}
\Delta F_{\nu}=\Delta U_{\nu}-T\Delta S_{\nu},\label{eq:freeE}\end{equation}
 which includes an entropic contribution from removing a single ion
from solution and localizing it in the pore region. This entropic
contribution is $\Delta S_{\nu}=k_{B}\ln\left(V_{p}n_{0}\right)$,
where we have assumed an ideal ionic solution and $V_{p}$ is the
volume of the pore and $n_{0}$ is the bulk salt concentration for
all species $\nu$. The free energy change is plotted in Figs. \ref{fig:EandImono}
and \ref{fig:EandIdi} versus the effective pore radius and it is
substantial when the latter becomes smaller than the outer hydration
layer.

\section{Ionic Currents\label{sec:Ionic-Currents}}

We now want to relate these energy barriers to the ionic current through
the pore %
\footnote{The most detailed information regarding ion channels and physical
processes in nanopores is provided by Molecular Dynamics (MD) - but
MD simulations are not able to reach the necessary time scales required
to extract the full information on the current. Indeed, there is a
hierarchy of approaches going down from macroscopic to microscopic
models: continuum models - Poisson-Boltzmann, Poisson-Nernst-Planck;
Brownian dynamics; classical then quantum Molecular Dynamics. In practice,
some combination of the different approaches is often used, such as
calculating structural/energetic properties from MD and using them
to construct simpler model systems that can then be tested experimentally.
This is the approach we have followed in this work. %
}. We do this by solving the Nernst-Planck equation in one dimension.
Since this model consistently solves for both drift and diffusion
contributions to ionic transport, and yields a compact analytical
expression, we use it below with the energetic barriers found from
the above model of dehydration. Even though this analytical model
does not include some effects such as ion-ion interaction, we expect
that it is qualitatively accurate as discussed along with its derivation.

The steady-state Nernst-Planck equation (see, e.g., \cite{Neumcke69-1,Eisenberg95-1,Hille01-1})
for species $\nu$ in one dimension (assuming variability on the pore
cross-section is not important) is \begin{equation}
J_{\nu}=-q_{\nu}D_{\nu}\left[\der{n_{\nu}\left(z\right)}z+\frac{q_{\nu}}{k_{B}T}n_{\nu}\left(z\right)\der{\Phi_{\nu}\left(z\right)}z\right],\label{eq:NP}\end{equation}
 where $J_{\nu}$ is the charge flux for species $\nu$, $z\in\left(0,l\right)$
is the axial coordinate along the pore axis of length $l$, $n_{\nu}\left(z\right)$
is the ion density, $D_{\nu}$ is the diffusion coefficient (assumed
to be position independent), and $\Phi_{\nu}\left(z\right)$ is the
position-dependent potential (including both electrostatic and other
interactions that change the energy within the pore). A full solution
would require solving the density and potential within the reservoirs
and pore simultaneously (see, e.g., Ref. \cite{Luchinsky09-1}). However,
we deal with high-resistance pores. Thus, we approximate the left
($L$) and right ($R$) reservoirs with constant concentrations $n_{L}$
and $n_{R}$, and the boundary conditions at the edge of the pore
are $n_{\nu}\left(0\right)=n_{L}$ and $n_{\nu}\left(l\right)=n_{R}$.
This is equivalent to assuming that as soon as an ion leaves or enters
the pore, the ions in the immediate surroundings of the pore equilibrate
rapidly to their prior distributions. Thus, multiplying by $\exp\left(q_{\nu}\Phi_{\nu}\left(z\right)/k_{B}T\right)$
to get \[
J_{\nu}e^{q_{\nu}\Phi_{\nu}\left(z\right)/k_{B}T}=-q_{\nu}D_{\nu}\der{}z\left[n_{\nu}\left(z\right)e^{q_{\nu}\Phi_{\nu}\left(z\right)/k_{B}T}\right]\]
 and integrating yields the flux for species $\nu$ as \[
J_{\nu}=-q_{\nu}D_{\nu}\frac{n_{R}e^{q_{\nu}\Phi_{\nu}\left(l\right)/k_{B}T}-n_{L}e^{q_{\nu}\Phi_{\nu}\left(0\right)/k_{B}T}}{\int_{0}^{l}dze^{q_{\nu}\Phi_{\nu}\left(z\right)/k_{B}T}}.\]
 We make the further simplifying assumption that the electrostatic
potential drops linearly over the pore - recognizing that in the presence
of a significant potential barrier, e.g., due to the stripping of
the water molecules from the hydration layers and in the absence of
surface/fixed charges in the pore, the ionic density in the pore is
small and thus the field is due to ionic charge layers on both sides
of the pore. Results from many works that include ion-ion interactions
indeed find a linear drop of the potential across the pore (see, e.g.,
Ref. \cite{Krems10-1}). In this case, ions form a capacitor across
the pore and every so often one ion translocates through the pore.
The {}``healing'' time for the loss of this ion is very short~\cite{Krems10-1}
and, thus, the field (potential drop) is not strongly affected %
\footnote{One may worry that these charge layers - which mainly are due to excess
ionic density - invalidate the assumption of constant ionic density
outside the pore. A quick estimate of the excess density comes from
the surface charge (for two parallel plates) necessary to give a typical
potential drop of 100 mV over $100\,\textrm{\AA}$. This is $\sigma=\epsilon_{0}V/l\approx5.5\times10^{-6}\, e/\,\textrm{\AA}^{2}$.
This surface charge is likely contained in a layer $\sim10\,\textrm{\AA}$
thick, giving a density $5.5\times10^{-7}\, e/\,\textrm{\AA}^{3}$.
For comparison, at 1 M concentration the density is $6\times10^{-4}\, e/\,\textrm{\AA}^{3}$
- that is, orders of magnitude larger than the variation in density
necessary to give the electric field over the pore. However, increasing
the bias or decreasing the bulk concentration or pore length may invalidate
this assumption. %
}. Also, we assume that the potential barrier due to changes in these
other interactions is constant over the pore - this ignores a region
near the pore entrance, but will not qualitatively change the solution.
Therefore, the potential for species $\nu$ can be written as\[
\Phi_{\nu}\left(z\right)=z\frac{V}{l}+\frac{\Delta F_{\nu}}{q_{\nu}}\]
 when $z\in\left(0,l\right)$. The boundaries are given by $\Phi_{z}\left(0\right)=0$
and $\Phi_{z}\left(l\right)=V$. Performing the remaining integral
and for equal reservoir densities (our case), $n_{L}=n_{R}=n_{0}$,
we get \[
J_{\nu}=-\frac{q_{\nu}^{2}n_{0}D_{\nu}V}{lk_{B}T}e^{-\Delta F_{\nu}/k_{B}T}.\]
 Relating the diffusion coefficient to the mobility via the Einstein
relation, $\mu_{\nu}=q_{\nu}D_{\nu}/k_{B}T$ , and putting in the
constant electric field $E=V/l$, one obtains \begin{equation}
J_{\nu}=-q_{\nu}n_{0}\mu_{\nu}Ee^{-\Delta F_{\nu}/k_{B}T}.\label{eq:flux}\end{equation}
 That is, the flux of an ionic species is proportional to the electric
field and density, where the latter is suppressed by a Boltzmann factor
\cite{Zwolak09-1}.

Now that we have an expression relating the energy barrier to the
transport properties, we can calculate the current as a function of
effective pore radius by multiplying Eq. \eqref{eq:flux} by the area
of the pore to get \begin{align}
I_{\nu} & =2\pi R_{p}^{2}J_{\nu}\nonumber \\
 & \equiv I_{\nu0}e^{-\Delta F_{\nu}/k_{B}T},\label{totcurr}\end{align}
 where we have defined a standard current $I_{\nu0}=-q_{\nu}n_{0}2\pi R_{p}^{2}\mu_{\nu}E$
that would flow in the absence of an energy barrier. The current~(\ref{totcurr}),
with the mobilities and energies in Table \ref{tab:values}, along
with Eqs. \eqref{eq:LayerU}-\eqref{eq:freeE}, is plotted in Figs.
\ref{fig:EandImono} and \ref{fig:EandIdi} as a function of effective
pore radius and for a field of $1\,\mathrm{mV/\textrm{\AA}}$ %
\footnote{We note that for all layers to be present, the applied field can not
be stronger than the ion's field of $\sim0.3\,\mathrm{V}/\textrm{\AA}$
- the magnitude of a monovalent ion's field within the third layer
($\sim7\textrm{\AA}$ from the ion) - and approximately double that
for divalent ions. In this way, the hydration layer structure will
not be significantly perturbed.%
}. The energetic barriers create sudden drops when the pore radii are
congruent with a hydration layer radius. These correspond to the quantized
steps in the conductance.

\section{Effect of noise\label{sec:Noise}}

In a real experiment, there will also be fluctuations in the energetic
barrier due to the fact that the hydration layers are not defined
by their time-averaged value (i.e., they are not perfect spherical
shells) and also due to fluctuations of the water structure and contents
of the pore (both within a single experiment and also structural variations
between experiments). Thus, we also examine the effect of these fluctuations
and the current noise they induce. Thus, we calculate an averaged
current for species $\nu$ as \begin{equation}
\avg{I_{\nu}}=\avg{I_{\nu0}e^{-\Delta F_{\nu}/k_{B}T}}.\label{eq:AvgI}\end{equation}
 We consider two specific models: Gaussian fluctuations of the free
energy with a standard deviation proportional to the free energy barrier
at a fixed pore radius and Gaussian fluctuations in the effective
pore radius. The latter was also considered previously \cite{Zwolak09-1}
where it was found that this type of noise smooths out the visibility
of the drops in conductance (i.e., the peaks in the derivative $d\avg{I_{\nu}}/dR_{p}$
become smoother with increasing noise). However, it was also shown
that this fluctuation induces a peak in the relative current noise
that is much less sensitive to the strength of the fluctuations -
thus giving an alternative method to detect the effect of the hydration
layers. We develop a model for this relative noise here but do not
perform the calculation of Eq. \eqref{eq:AvgI} for all the different
species.

\emph{Fluctuating energy barrier} - The first model we consider is
an energy barrier that fluctuates according to a Gaussian distribution.
We neglect fluctuations that make the barrier negative, so that the
average current is \begin{equation}
\avg{I_{\nu}}=\frac{I_{\nu0}}{\mathcal{N}_{\sigma}}\int_{0}^{\infty}d\left(\Delta F\right)\, e^{-\Delta F/k_{B}T}e^{-(\Delta F-\Delta F_{\nu})^{2}/2\sigma^{2}},\label{eq:EnergyFluct}\end{equation}
 where $\sigma$ is the standard deviation of the fluctuations and
\[
\mathcal{N}_{\sigma}=\int_{0}^{\infty}d\left(\Delta F\right)e^{-(\Delta F-\Delta F_{\nu})^{2}/2\sigma^{2}}\]
 is the normalization. The average current is thus \[
\avg{I_{\nu}}=I_{\nu0}Ae^{-\left(\Delta F_{\nu}-\sigma^{2}/2k_{B}T\right)/k_{B}T},\]
 where the factor $A$ is \[
A=\frac{\erfc\left[\left(-\Delta F_{\nu}+\sigma^{2}/k_{B}T\right)/\sqrt{2\sigma^{2}}\right]}{\erfc\left[-\Delta F_{\nu}/\sqrt{2\sigma^{2}}\right]}.\]

The value of $A$ for small $\sigma$ is very close to $1$. Thus,
the effect of a fluctuating energy barrier with small fluctuations
is simply to lower the energy barrier by an amount $\sigma^{2}/2k_{B}T$.
For stronger fluctuations, the factor containing the complementary
error function, $\erfc$, gives different limiting dependencies of
the average current as the fluctuation strength $\sigma$ is increased.
However, large fluctuations are well outside the realm of validity
of the present model. 

The relative noise in the current provides even more information.
The relative noise is \[
\dI=\frac{\sqrt{\avg{I^{2}}-\avg I^{2}}}{\avg I}.\]
 The expectation value of the square of the current is given by \begin{align}
\avg{I_{\nu}^{2}} & =\frac{I_{\nu0}}{\mathcal{N}_{\sigma}}\int_{0}^{\infty}d\left(\Delta F\right)\, e^{-\Delta F/k_{B}T}e^{-(\Delta F-\Delta F_{\nu})^{2}/2\sigma^{2}}\nonumber \\
 & =BI_{\nu0}^{2}e^{-\left(2\Delta F_{\nu}-2\sigma^{2}/k_{B}T\right)/k_{B}T}.\end{align}
 Where the normalization is as before and the factor $B$ is given
by \[
B=\frac{\erfc\left[\left(-\Delta F_{\nu}+2\sigma^{2}/k_{B}T\right)/\sqrt{2\sigma^{2}}\right]}{\erfc\left[-\Delta F_{\nu}/\sqrt{2\sigma^{2}}\right]}.\]
 Thus, the relative current noise induced by an energy barrier with
fluctuations is \begin{equation}
\Delta I_{rel}=\sqrt{e^{\sigma^{2}/\left(k_{B}T\right)^{2}}\frac{B}{A^{2}}-1}.\label{eq:smallsig}\end{equation}
 For small fluctuations, $A$ and $B$ depend very weakly on $\sigma$
and are both very close to $1$, giving a relative current noise \[
\Delta I_{rel}\approx\sigma/k_{B}T.\]
 As expected, the relative noise increases with the strength of the
fluctuations. For fluctuations proportional to the energy barrier,
as shown in the Figs. \ref{fig:EandImono} and \ref{fig:EandIdi},
the fluctuations give rise to a monotonic increase in the relative
noise. Overall, the effect of fluctuations in the energy barrier is
to decrease the effective energy barrier and increase the current.
This reduces the magnitude of the drops in the conductance but does
not destroy their visibility. This would therefore help in observing
quantized ionic conductance. It is worth noting, however, that this
type of noise makes the step of the third hydration layer the most
pronounced. This seems an unlikely situation in actual experiments
and other types of noise need to be considered. 

\begin{figure}
\begin{centering}
\includegraphics[width=8cm]{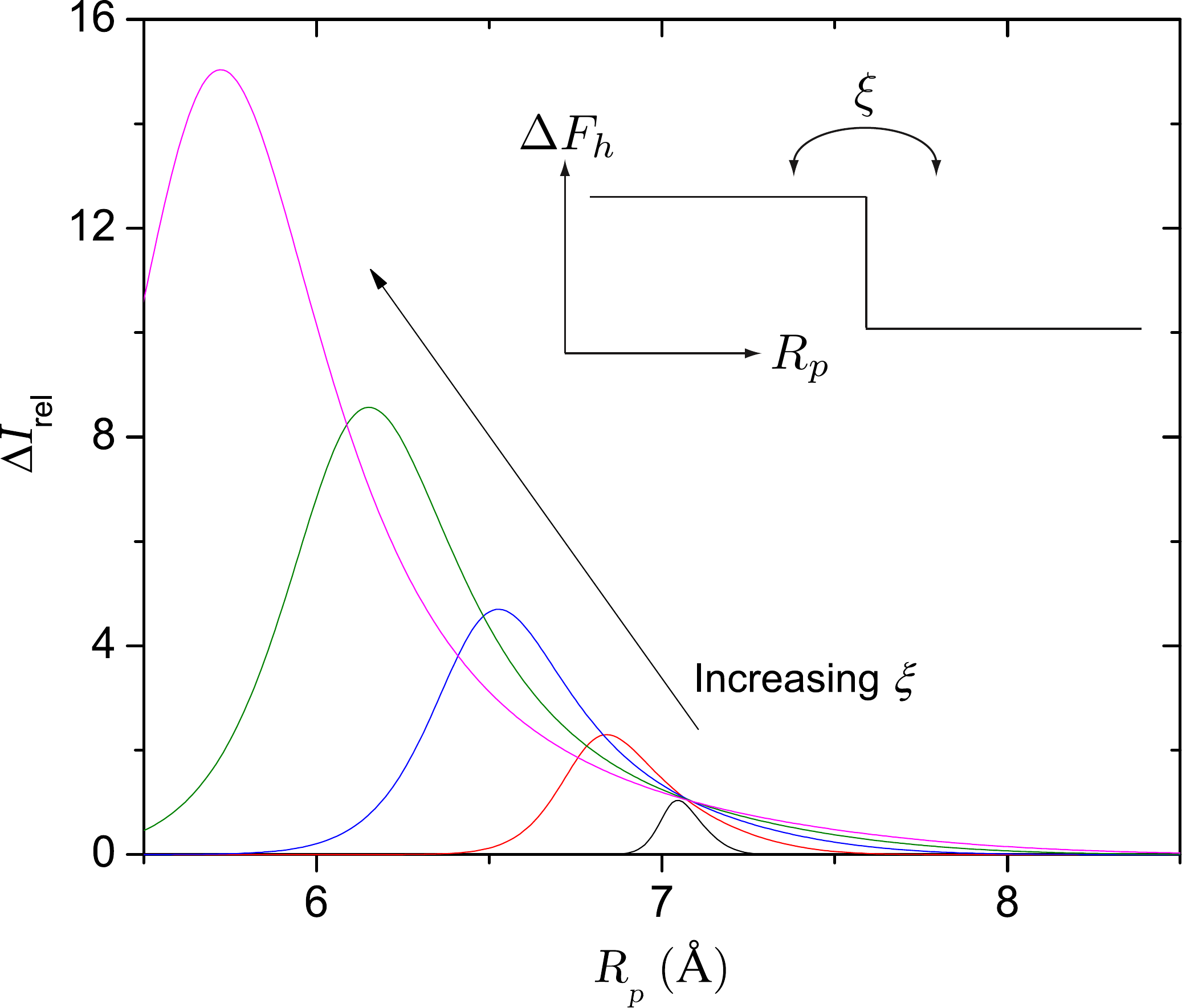} 
\par\end{centering}

\caption{Relative current noise induced by structural fluctuations in the effective
pore radius and/or hydration layer radius. The inset shows the approximate
change in free energy $\Delta F_{h}$ as a function of effective pore
radius in proximity to a hydration layer structure. The noise in the
pore radius induces fluctuations between the high and low energy states.
Here the third hydration layer radius of $\Cl$ is taken, $R_{h}=7.1\,\textrm{\AA}$.
See the text for details on $\Delta F_{h}$. The fluctuation strength
from right to left is $\xi=0.05,\,0.15,\,0.25,\,0.35,\,0.45\,\textrm{\AA}$.
\label{fig:dIrel}}

\end{figure}

\emph{Fluctuating effective pore radius} - In addition to the above
noise, one expects that there would be fluctuations in the radii of
the hydration layer/nanopore system. Previously, we demonstrated that
this type of noise can smear the effect of the steps in the current
\cite{Zwolak09-1}. As was seen, however, this noise also gives a
peak in the relative noise in the current that is much less sensitive
to the fluctuations than the average current. Here we develop a model
of this behavior by calculating the relative noise assuming fluctuations
across a single, perfect step in the free energy (see the inset of
Fig. \ref{fig:dIrel}).

The average current due to species $\nu$ when averaged over fluctuations
in the effective pore radius is \begin{equation}
\avg{I_{\nu}}=\frac{1}{\mathcal{N}_{\xi}}\int_{0}^{\infty}dR\, I_{\nu0}(R)e^{-\Delta F(R)/k_{B}T}e^{-(R-R_{p})^{2}/2\xi^{2}},\label{eq:RadAvg}\end{equation}
 where $ $$\xi$ is the standard deviation of the radial fluctuations,
$\mathcal{N}_{\xi}$ is the normalization, and the explicit $R$ dependence
has been included in both the barrier $\Delta F$ and the prefactor
$I_{\nu0}$. The dominant factor is the exponential of the free energy
barrier and the quadratic dependence of $I_{\nu0}$ on $R$ can be
ignored. For small fluctuations, the lower limit of the integral can
be extended to $-\infty$ and $\mathcal{N}_{\xi}\to\sqrt{2\pi\xi^{2}}$
to give \begin{equation}
\avg{I_{\nu}}=\frac{I_{\nu0}(R_{p})}{\sqrt{2\pi\xi^{2}}}\int_{-\infty}^{\infty}dR\, e^{-\Delta F(R)/k_{B}T}e^{-(R-R_{p})^{2}/2\xi^{2}}.\label{eq:RadAvgApp}\end{equation}
 Previously, we performed the averaging according to Eq. \eqref{eq:RadAvg}
\cite{Zwolak09-1}, but here we instead use Eq. \eqref{eq:RadAvgApp}
with the approximate energy barrier $\Delta F\left(R\right)=\Delta F_{h}\Theta\left(R_{h}-R\right)$
of a single hydration layer of radius $R_{h}$ and take $\bar{I}_{\nu0}$
to be the current in the absence of the barrier. The average current
then becomes \[
\avg{I_{\nu}}=\bar{I}_{\nu0}\left[e^{-\Delta F_{h}/k_{B}T}\left(1-C\right)+C\right],\]
 where \[
C=\frac{1}{2}\erfc\left(\frac{R_{h}-R_{p}}{\sqrt{2}\xi}\right).\]
 Similarly, for the square of the current one finds \[
\avg{I_{\nu}^{2}}=\bar{I}_{\nu0}^{2}\left[e^{-2\Delta F_{h}/k_{B}T}\left(1-C\right)+C\right].\]
 Although $\avg{I_{\nu}}$ and $\avg{I_{\nu}^{2}}$ are dependent
on the strength of the fluctuations, $\xi$, the relative current
noise has a universal behavior in the parameter $\tilde{R}=\left(R_{h}-R_{p}\right)/\sqrt{2}\xi$.
That is, all features in the relative noise would be present regardless
of the strength of the noise. However, the peak in the noise (see
below) shifts to smaller values of $R_{p}$ as the noise strength
is increased, which is qualitatively in agreement with the full averaging
(Eq. \eqref{eq:RadAvg}) performed in Ref. \cite{Zwolak09-1}.

The relative noise is \[
\dI=\frac{\sqrt{\left(1-e^{-\Delta F_{h}/k_{B}T}\right)^{2}C\left(1-C\right)}}{e^{-\Delta F_{h}/k_{B}T}\left(1-C\right)+C}.\]
 For large or small $R_{p}$, the relative noise goes to zero, which
can be seen from the properties of $\erfc$ that make $C\to1$ and
$C\to0$ for large and small $R_{p}$, respectively. In between these
limits, there would be nonzero relative noise, therefore indicating
that the relative noise would have a maximum. The peak in the relative
noise occurs for $R_{p}<R_{h}$. For a large energy barrier $\Delta F_{h}$,
this peak occurs when $C$ is small. Thus, we can approximate the
relative noise as \[
\dI\approx\frac{\sqrt{C}}{e^{-\Delta F_{h}/k_{B}T}+C}.\]
 This gives a peak in the noise when $C=e^{-\Delta F_{h}/k_{B}T}$
with a value \begin{equation}
\Delta I_{\mathrm{rel}}^{\star}\approx\frac{1}{2}e^{\Delta F_{h}/2k_{B}T}.\label{eq:dIpeak}\end{equation}
 The peak is exponentially large in the energy barrier. However, the
model with the electrostatic energy given by Eq. \eqref{eq:ElectroBarrier}
does not have an ideal step in the free energy (see, e.g., Figs.~\ref{fig:EandImono}
and~\ref{fig:EandIdi}). From previous work \cite{Zwolak09-1}, we
can identify the peaks, $R_{p}^{\star}$, and use $\Delta F_{h}\approx\Delta F_{\nu}\left(R_{p}^{\star}\right)$.
This is done in Fig. \ref{fig:dIrel} for $\Cl$. The model agrees
quantitatively with the full averaging performed in Ref. \cite{Zwolak09-1}.
The only feature missing is the additional background noise away from
the step due to the non-uniform energy barrier on both sides of the
step.


%

Thus, from this {}``two-channel'' model of noise we have found two
generic features: \textit{\emph{(}}\textit{i}\textit{\emph{)}} a peak
develops in the relative current noise that is exponentially high
with the hydration energy barrier, and (\textit{ii}\textit{\emph{)}}
it is present regardless of the noise strength, although its location
moves to smaller values of the pore radius with increasing noise (likewise,
the peak becomes wider). These features are in agreement with what
is found from performing the full averaging from Eq. \eqref{eq:RadAvg}
using the surface area model of the energy barrier \cite{Zwolak09-1}.
In the full model the fluctuations will eventually smooth out the
peak in the relative current noise. The latter, however, is still
much less sensitive than the average current drops, making the peak
in the relative current noise versus $R_{p}$ a robust indicator of
dehydration.

\begin{figure*}[t]
\begin{centering}
\includegraphics[width=15cm]{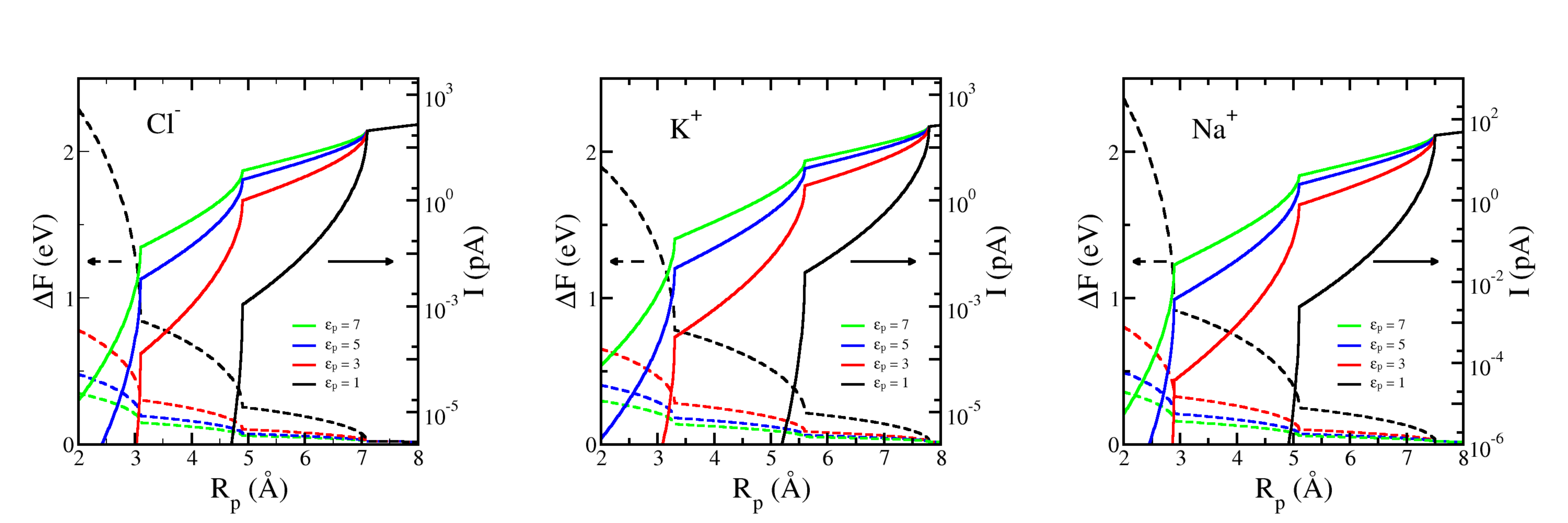} 
\par\end{centering}

\caption{Free energy changes, $\Delta F$, and currents versus the effective
pore radius for the monovalent ions and several values of $\epsilon_{p}$
($\epsilon_{p}=7,\,5,\,3,\,1$ from bottom to top) and for a field
of $1\,\mathrm{mV/\textrm{\AA}}$. The dashed lines indicate the free
energy change. The solid lines indicate the current (see text for
details). The currents are for $\epsilon_{p}=7,\,5,\,3,\,1$, from
top to bottom. \label{fig:DielectRedMono}}

\end{figure*}

\begin{figure*}[t]
\begin{centering}
\includegraphics[width=10cm]{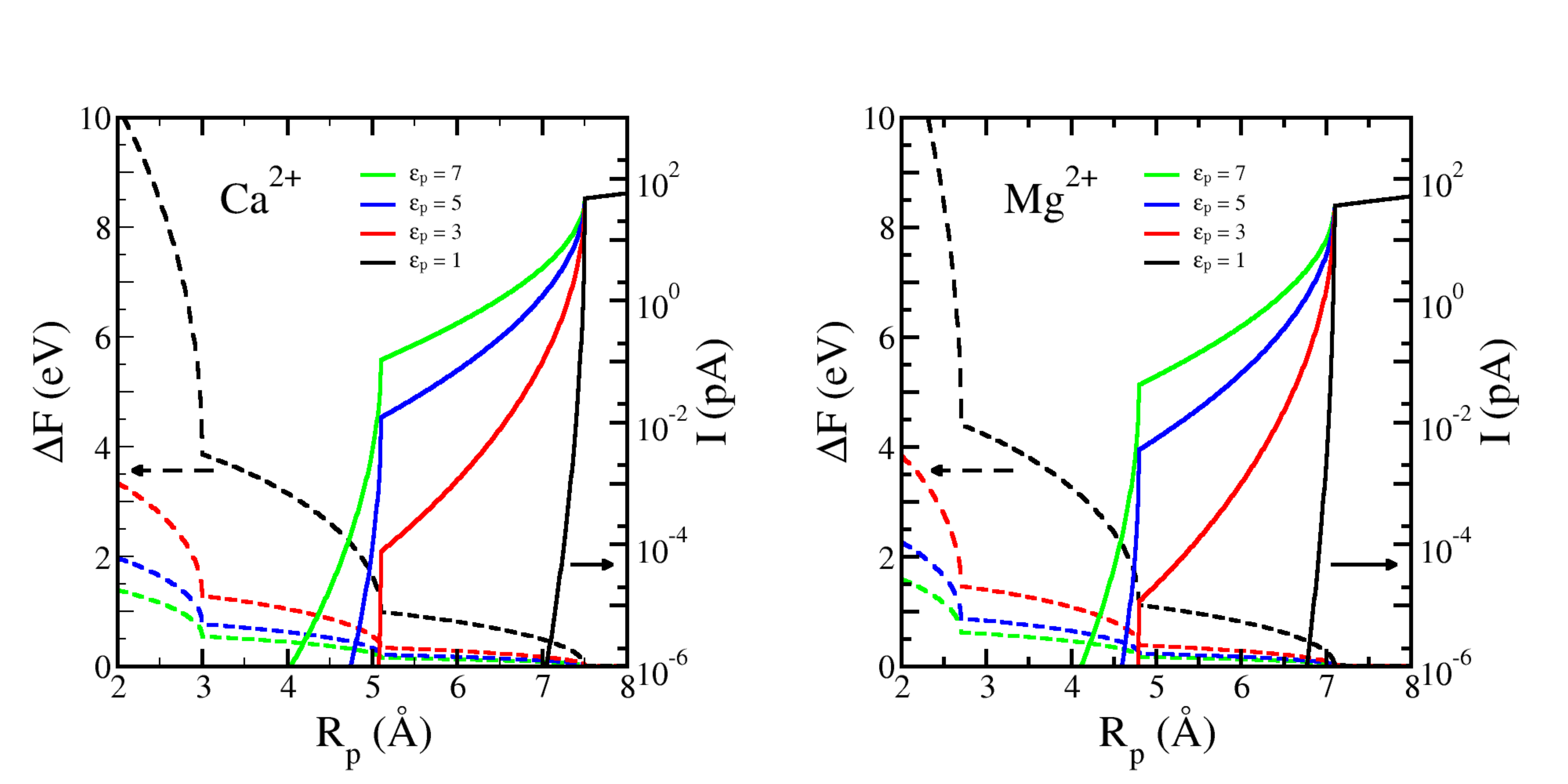} 
\par\end{centering}

\caption{Free energy changes, $\Delta F$, and currents versus the effective
pore radius for the divalent ions and several values of $\epsilon_{p}$
($\epsilon_{p}=7,\,5,\,3,\,1$ from bottom to top) and for a field
of $1\,\mathrm{mV/\textrm{\AA}}$.The dashed lines indicate the free
energy change. The solid lines indicate the current (see text for
details). The currents are for $\epsilon_{p}=7,\,5,\,3,\,1$, from
top to bottom. \label{fig:DielectRedDi}}

\end{figure*}

\emph{Barrier reduction} - In addition to the above fluctuations,
there are factors that reduce the energetic barrier, such as the presence
of some surface charge and/or dielectric screening in the pore. In
Eq. \eqref{eq:LayerU} we included a dielectric constant $\epsilon_{p}$
to represent a reduction in the hydration layer energy barrier from
these sources. We expect, however, that the introduction of this constant
overestimates the barrier reduction. It amounts to replacing the water
molecules screening the ion with a material of lower dielectric constant
but in the exact geometry of the water molecules. This is very unlikely
since the pore screening comes from the fixed surface of the pore
and thus in a different functional form. Nevertheless, it is instructive
to see how the drops in the current are reduced by this effective
lowering of the energy barrier. Figures \ref{fig:DielectRedMono}
and \ref{fig:DielectRedDi} show the energy barrier and current for
several values of this effective dielectric constant. We find that
even for fairly large $\epsilon_{p}$ ($\sim7$), the barriers are
large enough to give a noticeable drop in the current.

\emph{Bulk concentration} - We also mention the effect of changing
the concentration of ions in bulk. We have assumed that the hydration
layers are well formed away from the pore. Large ionic concentration
in bulk, however, would affect the formation of the hydration layers.
For a completely disassociated 1:1 salt, the ion-ion distance goes
as $\sim9.4/n_{0}^{1/3}\,\textrm{\AA}$ where the bulk concentration,
$n_{0}$, is given in mols/L. Thus, the inter-ion distance is $\sim9.4\,\textrm{\AA}$
for a 1 M solution, which is almost large enough to house both the
first and second hydration layers. However, concentrations lower than
1 M are preferable.

\emph{Some remarks} - We have discussed many of the factors that will
affect the detection of quantized ionic conductance. The most ideal
experiment would be one with pores of well-controlled diameter and
with smooth surfaces. Likewise, a small (or no) amount of surface
charge and a low dielectric constant of the pore will make the effect
more pronounced (and the ability to gate a pore, e.g., made of a nanotube,
would help even more in understanding the energetics of transport).
Not having these factors under control greatly affects the transport
properties of the ions \cite{Chu2009-1}. Therefore, pores made of,
for instance, semiconducting nanotubes may be ideal. Indeed, pores
made of these materials have been recently demonstrated \cite{Liu2010-1}.
However, rough surfaces that are present in pores made of, e.g., silicon
nitride, should still allow for quantized conductance to be observed,
so long as the variation of the effective radius of the pore is not
too strong. The noise in the effective radius of the pore was investigated
previously in Ref. \cite{Zwolak09-1}, where we found that only beyond
variation in the radius of $0.2-0.3\,\textrm{\AA}$ will the effect
be washed out. However, even beyond this variation magnitude, the
relative current noise signifies the presence of steps in the energy
barrier, thus giving a more robust indicator of the hydration layers'
effect on transport.

\section{Conclusions\label{sec:Conclusions}}

Ionic transport in nanopores is a fascinating subject with a long
history and impact in many areas of science and technology. Recent
work on developing aqueous-based nanotechnology and understanding
biological ion channels requires a firm understanding of how water
and ions behave in confined geometries and under non-equilibrium conditions
due to applied fields. For example, the quest for ultra-fast, single-molecule
DNA sequencing has yielded a number of proposals based on nanopores
\cite{Zwolak08-1}. Among them, transverse electronic transport \cite{Zwolak05-1,Lagerqvist06-1}
(whose theoretical basis includes the investigation of atomistic fluctuations
\cite{Zwolak05-1,Lagerqvist06-1,Lagerqvist07-1,Lagerqvist07-2} and
electronic noise in liquid environments \cite{Krems09-1}) and ionic
blockade \cite{Kasianowicz1996-1,Akeson1999-1,Deamer2000-1,Vercoutere2001-1,Deamer2002-1,Vercoutere2002-1,Vercoutere2003-1,Winters-Hilt2003-1}
have yielded promising recent experiments (Refs. \cite{Tsutsui10-1,Chang10-1}
and \cite{Clarke09-1,Stoddart09-1}, respectively). In all these cases,
both water and ions are present and will have a significant impact
on the signals and noise observed.

In this work, we have analyzed in detail the recent prediction of
quantized ionic conductance \cite{Zwolak09-1} and examined how different
aspects of the ion-nanopore system influence the detection of this
phenomenon. Namely, we have shown that the ion type affects very little
the radii at which the conduction should drop. High valency ions,
however, should give even more pronounced drops in the current and
thus may help in detecting this effect. Further, the presence of the
hydration layers gives a peak in the relative noise at pore radii
congruent with a layer radius. This relative noise is much less sensitive
to fluctuations than the average current, and provides a promising
approach to detect the effect of hydration.

Overall, quantized ionic conductance yields experimental predictions
that will shed light on the contribution of dehydration to ion transport
and we hope this work will motivate experiments in this direction. 
\begin{acknowledgments}
This research is supported by the U. S. Department of Energy through
the LANL/LDRD Program (M. Z.) and by the NIH-NHGRI (J. W. and M. D.). 
\end{acknowledgments}
\bibliographystyle{apsrev4-1} \bibliographystyle{apsrev4-1} \bibliographystyle{apsrev4-1}
\bibliography{DNAandPores}

\end{document}